\newif\ifDC \DCtrue
\ifDC
\documentclass[twocolumn,showpacs,preprintnumbers,amsmath,amssymb,prb,color]{revtex4}
\else
\documentclass[preprint,showpacs,preprintnumbers,amsmath,amssymb,prb]{revtex4}
\fi

\newif\ifSEC \SECtrue

\newif\ifTEST \TESTfalse

\usepackage[dvipdfm]{graphicx}
\usepackage{dcolumn}
\usepackage{bm}
\ifTEST
\usepackage[usenames]{color}
\definecolor{red}{cmyk}{0,1,1,0}

\fi

\newif\ifYBCO \YBCOfalse
\newcommand{\YBCO}{\ifYBCO{YBCO }\else{YBa$_2$Cu$_3$O$_y$ (YBCO)}\YBCOtrue\fi}
\newcommand{\YBCOE}{\ifYBCO{YBCO}\else{YBa$_2$Cu$_3$O$_y$ (YBCO)}\YBCOtrue\fi}
\newif\ifLBCO \LBCOfalse

\newif\ifLSCO \LSCOfalse
\newcommand{\LSCO}{\ifLSCO{LSCO }\else{La$_{2-x}$Sr$_x$CuO$_4$ (LSCO)}\LSCOtrue\fi}

\newif\ifLNSCO \LNSCOfalse

\newif\ifNCCO \NCCOfalse
\newcommand{\NCCO}{\ifNCCO{NCCO }\else{Nd$_{2-x}$Ce$_x$CuO$_4$ (NCCO)}\NCCOtrue\fi}

\newcommand{\RH}{$R_{\rm H}$ }

\newcommand{\rhoab}{$\rho_{ab}$ }
\newcommand{\Sab}{$S_{ab}$ }

\ifTEST
\newcommand{\citeAndoHTH}{{\red [Ref. AndoHTH]\cite{AndoHTH}}}
\newcommand{\citeAndoCarrierConc}{{\red [Ref. AndoCarrierConc]\cite{AndoCarrierConc}}}\newcommand{\citeSegawaHall}{{\red [Ref. SegawaHall]\cite{SegawaHall}}}
\else
\newcommand{\citeAndoHTH}{[Ref. 15]}
\newcommand{\citeAndoCarrierConc}{[Ref. 16]}
\newcommand{\citeSegawaHall}{[Ref. 13]}
\fi

\begin{document}

\title{Doping $n$-type carriers by La-substitution for Ba\\ in YBa$_2$Cu$_3$O$_y$ system}

\author{Kouji Segawa and Yoichi Ando}
\affiliation{Central Research Institute of Electric Power 
Industry, Komae, Tokyo 201-8511, Japan}

\date{\today}
\begin{abstract}
Thus far, there is no cuprate system where both $n$-type and $p$-type
charge carriers can
be doped without changing the crystallographic structure.
For studying the electron-hole symmetry in an identical structure,
we try to dope $n$-type carriers to YBa$_2$Cu$_3$O$_y$
system by reducing oxygen content
and substituting La$^{3+}$ ions for Ba$^{2+}$.
Single crystals of La-doped YBa$_2$Cu$_3$O$_y$ are grown by a flux method
with Y$_2$O$_3$ crucibles and it is confirmed that
La actually substitutes $\sim$13\% of Ba.
The oxygen content $y$ can be varied between 6.21 and 6.95
by annealing the crystals in an atmosphere with controlled
oxygen partial pressure.
The in-plane resistivity $\rho_{ab}$ at room temperature
was found to increase with decreasing oxygen content $y$ down to 6.32,
but interestingly further decrease in $y$ results in a {\it decrease} in $\rho_{ab}$.
The most reduced samples with $y=6.21$ show
$\rho_{ab}$ of $\sim$ 30 m$\Omega$ cm at room temperature,
which is as much as seven orders of magnitude smaller than the maximum value
at $y=6.32$.
Furthermore,
both the Hall coefficient and the Seebeck coefficient of the $y=6.21$ samples
are found to be negative at room temperatures.
The present results demonstrate that the non-doped
Mott-insulating state has been crossed upon reducing $y$
and $n$-type carriers are successfully doped in this material.   

\end{abstract}


\maketitle

\newlength{\figurewidth}
\newlength{\figurewidthnarrow}
\ifDC
\setlength{\figurewidth}{85mm}
\setlength{\figurewidthnarrow}{65mm}
\else
\setlength{\figurewidth}{135mm}
\setlength{\figurewidthnarrow}{100mm}
\fi



High-$T_{\rm c}$ superconducting cuprates are classified
into two types in terms of the sign of charge carriers: one has
$p$-type charge carriers
(hole-doped system) and the other has $n$-type ones
(electron-doped system) \cite{TokuraNature89}.
The magnetic and superconducting
phase diagrams of these types are known to be
different from each other, and thus
a whole phase diagram is conventionally drawn by connecting
those at the non-doping point.
However, discontinuity remains in the low-carrier limit, because
there has been no prototypical system in which the type of carriers
can be continuously changed from $p$-type to $n$-type while keeping
an identical crystallographic structure.
In the so-called ``214"-type layered cuprates either electrons or holes can be doped;
however, hole-doped 214 systems, for example \LSCO,
have $T$- or $T^\ast$-structure, whereas electron-doped 214 systems
like \NCCO\, have different $T^\prime$-structure \cite{TokuraPRB89}.
For the study of electron-hole symmetry in cuprates,
it would be very useful if one could change the sign of charge carriers without
changing the crystallographic structure unlike the case of 214 systems.

After the failure of various efforts to synthesize electron-doped cuprates,
an empirical relation between the doping and
crystallographic structures was proposed \cite{TokuraJJAP90}.
The coordination number of Cu atoms seems important for
determining the sign of carriers;
$p$-type carriers are favored with the coordination number of 5 or 6,
while $n$-type ones are favored for 4-coordinated Cu.
Indeed,
it seems difficult to dope electrons
to \LSCO with $T$-structure \cite{TsukadaPC05},
or to dope holes to an infinite-layer system with 4-coordinated Cu atoms
\cite{KuboPRB94}.
Furthermore, calculations of the Madelung potential \cite{OhtaPRB91} are consistent with the 
behavior.

In \YBCO,
the oxygen content can be varied in a very wide range
from hole-doped compositions to a non-doped insulating one.
The doping can be varied also by chemical substitutions
\cite{TokuraPRB88,TokiwaJJAP88};
for example,
it is well known that
substitutions of Ca$^{2+}$ for Y$^{3+}$ in the \YBCO system
increases the number of holes and shifts the doping range
into overdoped regime \cite{TallonPRB95}.
On the other hand,
La$^{3+}$-substitution for Ba$^{2+}$
is expected to decrease holes.
If the La-substitution for Ba is combined with
the reduction of oxygen,
the available doping range would shift
{\it across} the non-doping point.
This may allow one to study both the electron-doped and the hole-doped
regimes in the same system, and
the above empirical relation could be scrutinized.

In this work,
we explore the ways to change the sign of charge carriers
by combining the La substitution for Ba and the oxygen reduction
in the YBCO system.
In samples where La is substituted for 13\% of Ba,
we observe that the in-plane resistivity at room temperature
becomes maximum at $y\sim 6.32$,
and further reduction to $y=6.21$ leads to a decrease in the resistivity
down to $\sim 30 \,{\rm m}\Omega$ cm,
which is seven orders of magnitude smaller than the maximum value.
Furthermore, these $y=6.21$ samples show both a negative Hall coefficient
and a negative thermoelectric power, indicating that
electrons are successfully doped.

Single crystals of La-doped YBCO are grown by a flux method
using Y$_2$O$_3$ crucibles \cite{SegawaZndoped}.
In the starting material,
La$_2$O$_3$ is substituted for BaO$_2$ by 10\%,
and Y$_2$O$_3$ is provided by the crucibles.
The purity of the raw material is 99.99\% for CuO and 99.9\% for
BaO$_2$, La$_2$O$_3$ and Y$_2$O$_3$ (crucibles).
The actual composition of the grown crystals is analyzed by
the inductively coupled plasma atomic-emission 
spectroscopy (ICP-AES)
and is found to be Y$_{0.38}$Ba$_{1.74}$La$_{0.88}$Cu$_3$O$_y$.
Hence, La is substituted for both 13\% of Ba and 62\% of Y,
so we express the composition as
Y$_{1-z}$La$_z$(Ba$_{1-x}$La$_x$)$_2$Cu$_3$O$_y$
with $x = 0.13$ and $z = 0.62$.
Since Y and La ions have the identical valence number of +3,
the carrier density should not depend on $z$.
In this paper, we denote the specific composition
of our single crystals of
Y$_{0.38}$La$_{0.62}$(Ba$_{0.87}$La$_{0.13}$)$_2$Cu$_3$O$_y$
by YLBLCO$_y$
for simplicity.
The lattice parameters of as-grown crystals
are determined to be $a$=3.901 \AA\, and $c$=11.763 \AA\,
by the X-ray diffraction,
where, interestingly, $a$ is much longer than pristine \YBCOE.
In addition, no signal of any other phases is observed.
The annealing is performed under various conditions shown in Table 1 by using
a home-made furnace which can control the gas flow with a precise
oxygen partial pressure.
It should be noted that the oxygen control in our YLBLCO crystals
is completely reversible between $y$ = 6.21--6.95.
Also, no structural transition, except for the ordinary orthorhombic-to-tetragonal
transition, was reported for samples with similar compositions \cite{LindemerPC93};
therefore, crystallographically our samples are essentially unchanged down to $y=6.21$.
The oxygen content is determined mainly
by iodometric titration for polycrystalline samples
with the identical composition to YLBLCO single crystals
annealed at the same time,
which yields an error of less than $\pm$0.02.
The titration result is corroborated by precision measurements
of the mass of single crystals.
Twinned structures are observed only in $y=6.95$ samples,
but no detwinning is performed.
The in-plane resistivity and the Hall coefficient are measured by the 6-probe method \cite{SegawaOverlap}, but
a 2-wire method is employed for measuring the resistivity
of samples with very high resistance.
At least 2 samples are measured for each composition
in order to check for reproducibility,
and the accuracy of the present result is $\sim$20\%
for the 6-probe measurements, and $\sim$50\%
for the 2-wire measurements.

\begin{table}[b]
\begin{center}
\begin{tabular}{ccccc}
\hline
$y$ & $T_{\rm anneal}$ & duration & atmosphere & $T_{\rm c}$ \\
& ($^{\circ}$C) & time &  & (K) \\ \hline
6.95 & 485 & $\ge 7$ days & 1 atm O$_2$ & 25 \\
6.78 & 550 & 40 hours & 0.06 atm O$_2$ & -- \\
6.43 & 550 & 40 hours & 2$\times$10$^{-4}$ atm O$_2$  & -- \\
6.32 & 700 & 12 hours & 2$\times$10$^{-5}$ atm O$_2$  & -- \\
6.29 & 750 & 12 hours & 2$\times$10$^{-5}$ atm O$_2$  & -- \\
6.21 & 890 & 6 hours & 2$\times$10$^{-5}$ atm O$_2$  & -- \\ \hline
\label{Table1}
\end{tabular}
\end{center}
\caption{Annealing conditions for YLBLCO crystals.
$y$ is the measured oxygen content.}
\label{Annealing_conditions}
\end{table}

Figure 1(a) shows the temperature dependence of
the in-plane resistivity for YLBLCO with $y=6.95$
together with the data for pristine YBCO with $y=6.55$.
The zero-resistance $T_{\rm c}$ of the YLBLCO sample is
$\sim$25 K.
The temperature dependence of the Hall coefficient [Fig. 1(b)]
in YLBLCO$_{6.95}$ is found to be very close to
that in YBCO$_{6.55}$, and
this result suggests that
the hole concentration of YLBLCO$_{6.95}$
corresponds to that of YBCO$_{6.55}$,
which was suggested to be $\sim$7\% per Cu \citeSegawaHall.
Corroboratively, the slope of the temperature dependence 
of \rhoab in the YLBLCO$_{6.95}$
is similar to that in YBCO$_{6.55}$.
In YBCO system,
there is no simple relation between the oxygen content and
the exact hole concentration, because
the latter depends also on the oxygen ordering in the Cu-O chain layers
\cite{VealPRB90},
as well as on how the positive charge is transferred from the Cu-O chains
to the CuO$_2$ planes.
Therefore, it is not simple to understand how
the above two compositions produce
an identical hole concentration.
Nevertheless, the present result
might be helpful for estimating the hole concentration
of optimally-doped YBCO$_{6.95}$,
because the effect of the oxygen ordering is quite modest
in nearly fully-oxygenated samples.
As mentioned above, the hole concentration of YLBLCO$_{6.95}$ is
likely to be $\sim$7\%/Cu, and thus
in YBCO$_{6.95}$ it is inferred to be $\sim$20\%/Cu by simple calculation
(If so, the average valence of Cu in the Cu-O chain layers would be +2.50.).
We note that
the actual hole concentration in the CuO$_2$ planes can be
different from $\sim$20\%/Cu,
if the La-substitution affects the charge transfer from the chains
to the planes.

\begin{figure}
\begin{center}
\includegraphics[width=\figurewidthnarrow]{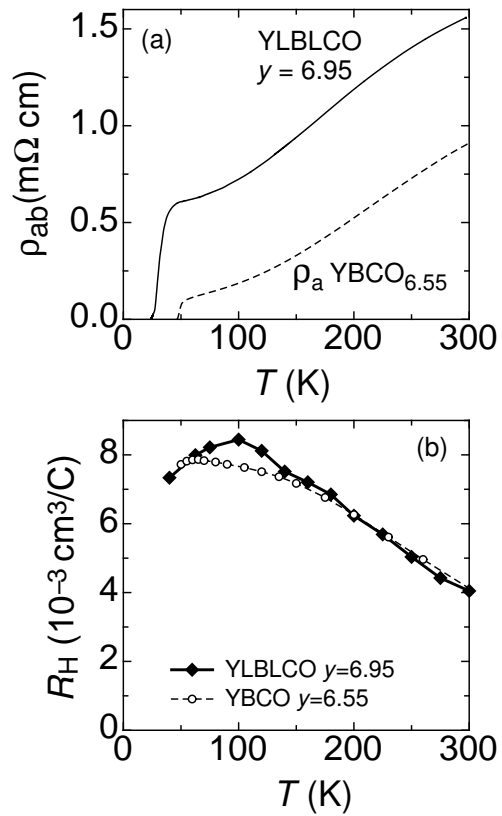}
\caption{The temperature dependences of
(a) the in-plane resistivity and
(b) the Hall coefficient
in YLBLCO with
$y=6.95$, together with those for
YBCO with $y=6.55$.}
\label{Fig1}
\end{center}
\end{figure}

The oxygen content of YLBLCO crystals can be decreased
by reduction annealings.
Figure 2 shows the evolution of the temperature dependence
of the in-plane resistivity upon changing $y$ in a semi-log plot.
\rhoab increases significantly with decreasing $y$
down to 6.32.
In samples with $y$=6.32,
\rhoab at room temperature becomes
as large as $\sim 10^5$ $\Omega$ cm,
which is about five orders of magnitude
larger than that of La$_2$CuO$_4$ \citeAndoHTH.
We emphasize that the observed insulating behavior is
not due to decompositions, because
the samples can be re-annealed to another composition
and give consistent results.
When the oxygen content is decreased further from 6.32,
interestingly, we observe a {\it decrease} in resistivity with decreasing $y$
[broken lines in Fig. 2].

\begin{figure}
\begin{center}
\includegraphics[width=\figurewidthnarrow]{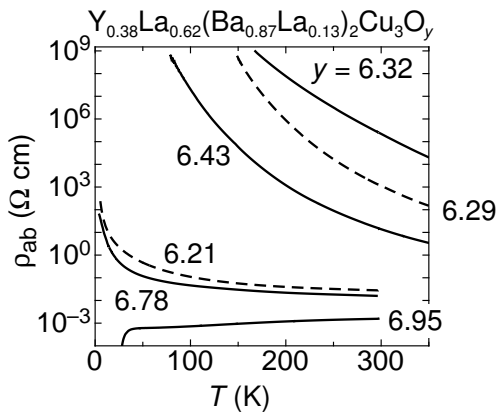}
\caption{Semi-log plot of the
temperature dependence of \rhoab
in YLBLCO with various oxygen contents.
The broken lines are for $y<$6.32, and
the solid lines are for larger $y$.}
\label{Fig_rho}
\end{center}
\end{figure}

Figures 3(a-c) show temperature dependences
of the transport properties
in the most reduced YLBLCO crystals with $y$=6.21.
The in-plane resistivity in YLBLCO$_{6.21}$
becomes 30 m$\Omega$ cm at room temperature
[Fig. 3(a)],
which is seven orders of magnitude smaller than that
in YLBLCO$_{6.32}$.
The temperature dependence of
\rhoab shows an insulating behavior below 300 K.
The Hall coefficient shown in Fig. 3(b) is negative
all the way below 300 K.
The temperature dependence of \RH is quite modest
except at low temperatures,
which gives confidence in estimating the carrier concentration
from the 300-K value of \RH to be $\sim$2\% per Cu \citeAndoCarrierConc.
Furthermore, the Seebeck coefficient also shows
a negative sign [Fig 3(c)]. These results clearly show that
the sign of the charge carriers in this sample is {\it negative}.
At the moment we cannot rule out the possibility that
the $n$-type carriers are doped to the Cu-O chain layers
rather than to the CuO$_2$ planes;
however, 
such a possibility is very unlikely
because the Hall mobility in YLBLCO$_{6.21}$, which is obtained to be
1.0 cm$^2$/Vs at 300 K, is comparable
to that in YLBLCO$_{6.95}$ (2.7 cm$^2$/Vs)
and pristine \YBCO \citeSegawaHall;
such a high mobility of $n$-type carriers would not be expected
in the Cu-O chain layers at $y$=6.21, at which the chains are
very disordered and fragmented.

\begin{figure}
\begin{center}
\includegraphics[width=\figurewidthnarrow]{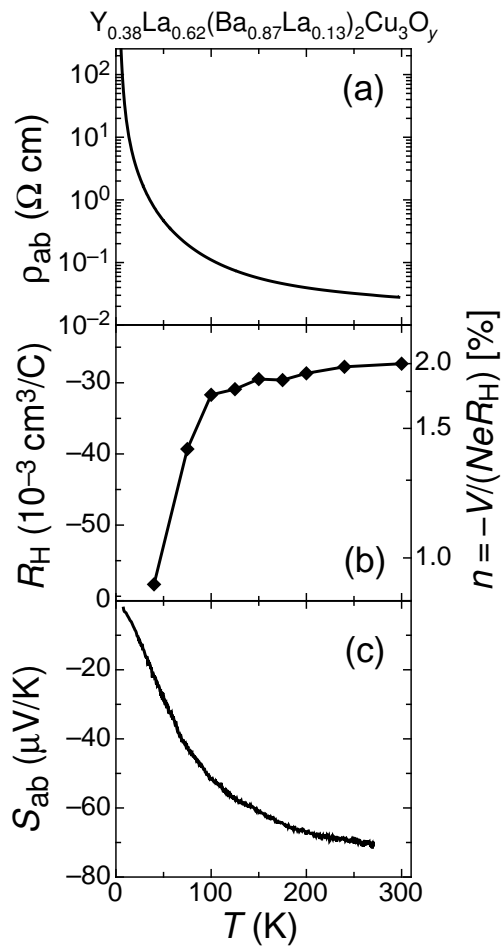}
\caption{The temperature dependences of three transport properties,
(a) the in-plane resistivity,
(b) the Hall coefficient, and (c) the Seebeck coefficient,
in YLBLCO with
$y=6.21$.}
\label{Fig_ED}
\end{center}
\end{figure}

Figures 4(a) and (b) show the $y$ dependences of \rhoab and \Sab
at fixed temperatures.
\rhoab shows a steep maximum at $y\sim 6.32$,
where apparently there are few carriers.
The Hall coefficient in this sample measured at room temperature 
is positive and $\sim 5\times$10$^4$ cm$^3$/C
within an error of $\sim$50\%,
which corresponds to the
hole concentration of $\sim 1\times 10^{-6}$ per Cu.
These observations indicate that
YLBLCO$_{6.32}$ is very close to the non-doped state.
The sign of \Sab correspondingly changes from positive to negative
with decreasing $y$ across $\sim 6.3$ [Fig. 4(b)].
The absolute value of \Sab appears to increase
upon approaching the non-doped composition.
Probably, \Sab for $y=6.32$ is in the middle of a jump from positive
to negative and happens to be intermediate.
Similar behavior of the Seebeck coefficient
is observed also in GdBaCo$_2$O$_{5+x}$ system,
where continuous ambipolar doping is possible \cite{TaskinPRB05}.

As discussed above, about 2\% of electrons per Cu
is successfully doped to \YBCO system, but
the system still remains insulating.
Doping more electrons to this system is desirable, but
unfortunately further reduction of YLBLCO
has not been successful yet
\cite{ED00}.
%
One may guess that
increasing $x$
is helpful for increasing $n$-type carriers.
In this respect, we have also grown crystals with larger $x$
(= 0.18 and 0.32), but
it turns out that
low enough $y$ cannot be achieved
in those high-$x$ samples.
Hence, there is apparently a delicate balance between
the $x$ value and the lowest achievable $y$ value.
This is probably the main reason why
this system has not been discovered as an electron-doped system.
There still remains a possibility that optimizing the $x$ value
allows further electron doping to make
the system metallic and/or superconducting.

\begin{figure}
\begin{center}
\includegraphics[width=\figurewidth]{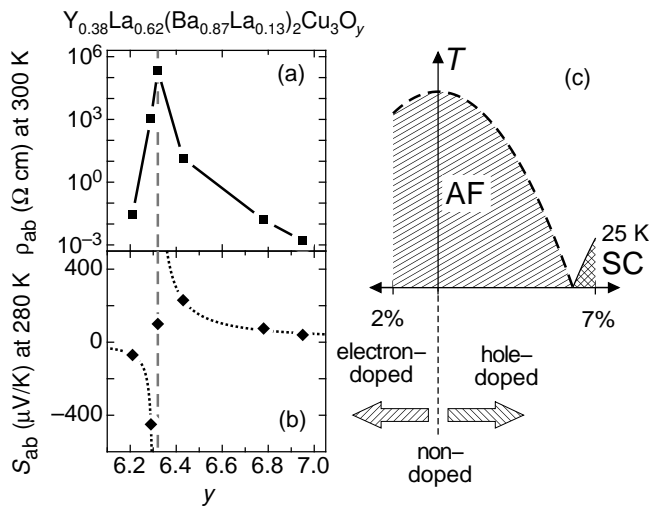}
\caption{Oxygen-content dependence of
(a) the in-plane resistivity at 300 K and (b) the Seebeck coefficient at 280 K
in YLBLCO.
(c) Speculative phase diagram of YLBLCO.
}
\label{Fig_PD}
\end{center}
\end{figure}

Figure 4(c) shows a speculative phase diagram
for YLBLCO.
In principle,
one would be able to
draw the phase diagram without
discontinuity at the non-doping point
just by determining
the N\'{e}el temperature $T_{\rm N}$ by a suitable means.
We would like to emphasize that,
to the best of our knowledge,
this system is the first bilayer cuprate having $n$-type carriers.

Since the Cu atoms in the CuO$_2$ planes in the so-called ``123" structure
are 5-coordnated,
it appears that the coordination number of Cu
does not play a prohibitive role for determining
the sign of charge carriers.
Instead, the Cu-O bond length in the CuO$_2$ planes
may be important, because
that length in the present system ($\simeq 1.95$ \AA)
is notably longer than that in other hole-doped cuprates
and is almost equal to the NCCO system.

In conclusion,
about 2\% of $n$-type carriers are successfully
doped in a Y-based bilayer cuprate
in which La is substituted for Ba by 13\%.
In this system
one can change the doping from $p$-type to
$n$-type by reducing the oxygen content.
Thus,
we have, for the first time,
spanned a doping range from negative to positive
across the non-doping state in the same crystal.
This opens a new avenue for studying the electron-hole
symmetry (asymmetry) in cuprates.

This research was supported by the Grant-in-Aid for Science
provided by the Japanese Society for the Promotion of Science.


\newpage

\ifTEST
\bibliography{../bib/segawa_HTS}
\else

\fi

\newpage

\end{document}